\documentclass[referee,10]{mn2e}
\usepackage{graphics}

\makeatletter

\providecommand{\LyX}{L\kern-.1667em\lower.25em\hbox{Y}\kern-.125emX\@}

\usepackage{graphicx}

\makeatother
\begin{document}

\title{Constraints on perfect fluid and scalar field dark energy models
from future redshift surveys }

\baselineskip 14pt

\author[L. Amendola, C. Quercellini \& E. Giallongo]{L. Amendola$^{1}$, C. Quercellini$^{1,2,3}$ \& E. Giallongo$^{1}$
\\$^{1}$INAF/Osservatorio Astronomico di Roma\\
Via Frascati 33, 00040 Monteporzio Catone (Roma), Italy\\
$^{2}$Universit\`a di Roma Tor Vergata, Dip. di Fisica, \\
Via della Ricerca Scientifica 1, 00133 Roma, Italy\\
$^{3}$University of Oxford, Keble Road, OX1 3RH, Oxford, UK}
\date{Accepted ....,
      Received }
\pagerange{\pageref{firstpage}--\pageref{lastpage}}
\maketitle
\label{firstpage}\date{\today }


\begin{abstract}
\baselineskip 14pt
We discuss the constraints that future photometric and spectroscopic
redshift surveys can put on dark energy through the baryon oscillations
of the power spectrum. We model the dark energy either with a perfect
fluid or a scalar field and take into account the information contained
in the linear growth function. We show that the growth function helps
to break the degeneracy in the dark energy parameters and reduce the
errors on $w_{0},w_{1}$ roughly by 30\% making more appealing multicolor
surveys based on photometric redshifts. We find that a 200 $\deg ^{2}$
spectroscopic survey reaching $z\approx 3$ can constrain $w_{0},w_{1}$ to within
$\Delta w_{0}=0.21,\Delta w_{1}=0.26$ and to $\Delta w_{0}=0.39,\Delta w_{1}=0.54$
using photometric redshifts with absolute uncertainty of 0.02. In
the scalar field case we show that the slope $n$ of the inverse power-law
potential for dark energy can be constrained to $\Delta n=0.26$ (spectroscopic redshifts) or $\Delta n=0.40$ (photometric redshifts), i.e. better than 
 with future ground-based supernovae surveys or CMB data.
\end{abstract}



\baselineskip 14pt
\section{Introduction}

Dark energy is the most elusive component of the universe: it is dark,
it does not cluster and, in most models, it is subdominant in cosmic
history until recently. On the other hand, it is also the most pervasive
component, in that it affects background expansion, linear dynamics
and possibly non-linear as well.

The characterization of DE has been so far based almost uniquely on
background tests at rather low redshifts ($z<1$: Riess et al. 1998,
Perlmutter et al. 1999, Tonry et al. 2003, Riess et al. 2004) or very
large redshifts ($z\approx 1000$: e.g. Netterfield et al. 2002, Halverson
et al. 2002, Lee et al. 2002, Bennett et al. 2003), although tests
at intermediate redshifts based on strong lensing have also appeared
(Soucail, Kneib \& Golse 2004). These tests are based essentially
on estimations of luminosity $D_{L}$ or angular-diameter distances
$D(z)$, i.e. on integrals of the Hubble function $H(z)$ which, in
turn, contains an integral of the equation of state . If we aim at making
real progress it is then necessary to cross this information with
probes at intermediate $z$ and with other observables such as the
linear growth function $G(z)$. The goal of this paper is to show
how future observations of large scale structure (LSS) at $z$ up
to 3 can set interesting limits on models of DE taking advantage of the complementarity
between $H,D$ and $G$.

The method we use is based on recent proposals (Linder 2003, Blake
\& Glazebrook 2003, Seo \& Eisenstein 2003) to exploit the baryon
oscillations in the power spectrum as a standard ruler calibrated
through CMB acoustic peaks. In particular, Blake \& Glazebrook (2003)
and Seo \& Eisenstein (2003; hereinafter SE) have shown the feasibility of large
(100 to 1000 square degrees) spectroscopic surveys at $z\approx 1$
and $z\approx 3$ to put stringent limits to the equation of state
$w(z)$ and its derivative. SE have also shown that the redshift error
introduced by a photometric survey (with a relative error on $1+z$
of $1\%$) could be compensated only by a 20-fold increase
of the sampled volume.

This paper has a threefold motivation. First, we extend the Fisher
matrix method by taking into account the information on the DE evolution
contained in the growth factor $G$ at various redshifts. While SE
marginalized over $G$, we find that its marked dependence on the
cosmological parameters, especially on $\Omega _{m}$, associated with its different
direction of degeneracy, are helpful in narrowing down the constraints.
Secondly, we extend the analysis to a scalar field model of DE with
inverse power-law potential (Ratra \& Peebles 1988), based on the
inverse power-law potential: this allows us to evaluate the growth
function without further assumptions and to exploit the properties
of tracking solutions (Steinhardt et al. 1999). Third, we pay special
attention to the possibility of testing DE with a photometric redshift
survey. Indeed the effect of the growth factor in reducing the confidence
regions of the cosmological parameters makes more appealing the use
of photometric redshift surveys to show any possible redshift evolution
of the DE. We explore the constraints derived on the cosmological
parameters from relatively wide multicolor surveys where the absolute
redshift uncertainty keeps almost constant around $\delta z=0.02-0.04$
in the redshift range $z=0-3.5$. Such surveys can be realized with
the current wide field imagers on 4-8m class telescopes.

Finally, it is interesting to speculate on another aspect of the method.
The baryon oscillation scale measures the angular-diameter distance
$D(z)$ while the SNIa method observes the luminosity distance. As
remarked recently by Bassett \& Kunz (2003) the two quantities are
related in a trivial way only in standard cosmologies: new physics
like photon decay can lead to an observable breaking of their {}``reciprocity''
relation. A photometric survey can be adapted to the scope since the
degradation of the error in $D(z)$ due to the redshift smearing is
relatively reduced (with respect to the error in $H(z)$), as it will
shown below.

\section{Fisher matrix method}

The method proposed in SE is based on the Fisher matrix, an approximation
to the likelihood function that provides under some conditions the
minimal errors that a given experiment may attain (Eisenstein, Hu
\& Tegmark 1999, EHT ). Here we review the method adopting the notation
of SE, to which we refer for further details. Our starting point is
a cosmological model that predicts the evolution of the Hubble parameter,
$H(z)$, and the growth of linear perturbations, $G(z)$. From $H(z)$
we can estimate the angular-diameter distance in flat space as ( we
put $c=1$)\begin{equation}
D(z)=(1+z)^{-1}\int _{0}^{z}\frac{dz'}{H(z')}\,.\end{equation}
 Let us choose a reference cosmology, denoted by a subscript $r$,
for instance $\Lambda CDM$, and calculate the matter power spectrum
in real space at $z=0$, $P_{0r}(k_{r})$. Then the predicted observable
galaxy power spectrum in a different cosmology $C$ (no subscript)
at $z$ in redshift space is\begin{equation}
P_{obs}(z;k_{r},\mu _{r})=P_{s}(z)+\frac{D_{r}^{2}(z)H(z)G^{2}(z)}{D^{2}(z)H_{r}(z)}\frac{\Omega _{m}^{1.2}(z)}{\beta ^{2}(z)}(1+\beta \mu ^{2})^{2}P_{0r}(k)\,,\end{equation}
 where $k$ is the wavenumber modulus and $\mu $ its direction cosine.
Several comments will clarify the meaning of this equation. $P_{s}(z)$
is a scale-independent offset due to incomplete removal of shot-noise.
The factor $(D^{2}/H)/(D_{r}^{2}/H_{r})$ takes into account the difference
in comoving volume between the two cosmologies. The factor $(1+\beta \mu ^{2})^{2}$
is the redshift distortion. $G(z)$ is the growth factor of the linear
density contrast $\delta _{m}$,\begin{equation}
G(z)=\delta _{m}(z)/\delta _{0}\,,\end{equation}
 where $\delta _{0}$ is the present density contrast (and therefore
$G(z)$ is normalized to unity today). The density parameter $\Omega _{m}(z)$
depends on the cosmology. In flat-space $\Lambda $CDM it is given
by \begin{equation}
\Omega _{m}(z)=\frac{\Omega _{m0}(1+z)^{3}}{\Omega _{m0}(1+z)^{3}+1-\Omega _{m0}}\,,\end{equation}
 and it is therefore parametrized by the present density parameter
$\Omega _{m0}$. The bias parameter $\beta (z)=\Omega _{m}^{0.6}(z)/b(z)$
(assumed scale-independent) is evaluated for the reference cosmology
using the formula\begin{equation}
\sigma _{8,g}=\beta (z)^{-1}\Omega _{m}^{0.6}(z)\sigma _{8,m}(z)\sqrt{1+\frac{2\beta (z)}{3}+\frac{\beta ^{2}(z)}{5}}\,, 
\end{equation}
 where $\sigma _{8}$ is the variance in spherical cells of 8 Mpc/$h$
for galaxies (subscript $g$) or matter ($m$) (notice that the variance
$\sigma _{8}$ is calculated using the power spectrum averaged over
$\mu $). Finally, we can include a redshift error by rescaling the
power spectrum\begin{equation}
P=P_{obs}e^{-k^{2}\mu ^{2}\sigma _{r}^{2}}\,,
\end{equation}
 where \begin{equation}
\sigma _{r}=(\delta z)/H(z)\,,
\end{equation}
 is the absolute error in distance and $\delta z$ the absolute error
in redshift.

We need also the relation between the line-of-sight and transverse
wavenumbers $k_{\bot },k_{\Vert }$ and the reference values: \begin{eqnarray}
k_{r\bot } & = & k_{\bot }D/D_{r}\,,\\
k_{r\Vert } & = & k_{\Vert }H_{r}/H\,.
\end{eqnarray}
 From these relations we derive the relation between the wavenumber
modulus $k$ and the direction cosine $\mu =\mathbf{k}\cdot \mathbf{r}/k$
in the reference cosmology and in the cosmology $C$\begin{eqnarray}
k & = & Rk_{r}\,,\\
\mu  & = & \frac{H\mu _{r}}{H_{r}R}\,,
\end{eqnarray}
 where \begin{equation}
R=\frac{\sqrt{D^{2}H^{2}\mu _{r}^{2}-D_{r}^{2}H_{r}^{2}(\mu _{r}^{2}-1)}}{DH_{r}}
\,.\end{equation}

The observed power spectrum in a given redshift bin $z_0$ depends therefore on a number of parameters,
denoted collectively $p_{i}$, such as $\Omega _{m0},G(z_0),H(z_0)$ etc,
as detailed in Table I. The redshift-dependent parameters ($P_s,H,G,D,\beta$ ) are assumed to be approximately constant in each redshift bin. Then we calculate, numerically or analytically,
the derivatives \begin{equation}
\left(\frac{\partial \log P}{\partial p_{i}}\right)_{ref}\,.\end{equation}
 Finally, the Fisher matrix is (Tegmark 1997)\begin{equation}
F_{ij}=\frac{1}{8\pi ^{2}}\int _{-1}^{1}\int _{k_{min}}^{k_{max}}\left(\frac{\partial \log P}{\partial p_{i}}\frac{\partial \log P}{\partial p_{j}}\right)_{ref}V_{eff}(k,\mu )k^{2}dkd\mu \,,\end{equation}
 where\begin{equation}
V_{eff}(k,\mu )=\left[\frac{nP(k,\mu )}{nP(k,\mu )+1}\right]^{2}V_{survey}\label{eq:stn}\,,\end{equation}
 $n$ being the number density of galaxies.

In the following we will use also the CMB Fisher matrix (Seljak 1996,
EHT)\begin{equation}
F_{ij,CMB}=\sum _{\ell }\sum _{X,Y}\frac{\partial C_{X,\ell }}{\partial p_{i}}(Cov_{\ell })_{XY}^{-1}\frac{\partial C_{Y,\ell }}{\partial p_{j}}\, ,\end{equation}
 where $C_{X,\ell }$ is the multipole spectrum of the component $X$,
and $X=T,E,B,C$ denote the temperature $T$, $E$ and $B$ polarization,
and the cross-correlation $TE$, respectively, with covariance matrix
$(Cov_{\ell })_{XY}$. We adopt the same specifications used in EHT
and in SE, corresponding to an experiment similar to the Planck satellite.
The spectra have been obtained with the Boltzmann code Cmbfast (Seljak
\& Zaldarriaga 1996). Once we have the Fisher matrices for all data
we sum to obtain the total combined matrix\begin{equation}
F_{ij}=F_{ij,CMB}+\sum F_{ij,survey}\, .\end{equation}
 The independent parameters to be included in the Fisher matrix are listed in
Table I.

\vspace{.2in}

\begin{center}\begin{tabular}{|c|c|c|}
\hline 
\multicolumn{3}{|c|}{Table I. Parameters}\\
\hline
\multicolumn{3}{|c|}{$P_{0}(k)$}\\
\hline
1&
 reduced total matter &
 $\omega _{m}\equiv \Omega _{m0}h^{2}$\\
\hline
2&
 reduced baryon dens.&
 $\omega _{b}\equiv \Omega _{b0}h^{2}$\\
\hline
3&
 optical thickness&
 $\tau $\\
\hline
4&
 primord. fluct. slope&
 $n_{s}$\\
\hline
5&
 total matter dens.&
 $\Omega _{m0}$\\
\hline
\multicolumn{3}{|c|}{For each survey at $z$}\\
\hline
6&
 shot noise&
 $P_{s}$\\
\hline
7&
 ang-diam. distance&
 $\log D$\\
\hline
8&
 Hubble param.&
 $\log H$\\
\hline
9&
 growth function&
 $\log G$\\
\hline
10&
 bias &
 $\log \beta $\\
\hline
\multicolumn{3}{|c|}{additional CMB parameters}\\
\hline
11&
 decoupling ang.-diam. dist.&
 $\log D$\\
\hline
12&
 tensor-to-scalar ratio&
 $T/S$\\
\hline
13&
 $C_{\ell }$ normalization&
 $\log A_{s}$ \\
\hline
\end{tabular}\end{center}

\vspace{.2in}

The main difference with respect to the method of SE is that we exploit
the information contained in the growth function. As it will be seen
below, this information is of great help in reducing the bounds on
the DE parameters, for two concurrent reasons: first, $G$ has a marked
dependence upon the cosmological parameters, i.e. $d\log G/d\Omega _{m0}$
is at $z>1$ as large as $d\log D/d\Omega _{m0}$ and quite larger
than $d\log H/d\Omega _{m0}$ (see Fig. 1); second, the parameter
degeneracy in $G(z)$ has a slightly different direction with respect
to the degeneracies in $H$ and $D$ (see the discussion in Cooray,
Huterer \& Baumann 2004 ). 

\begin{figure}
\includegraphics[  bb=0bp 400bp 598bp 843bp,
  clip,
  scale=0.7]{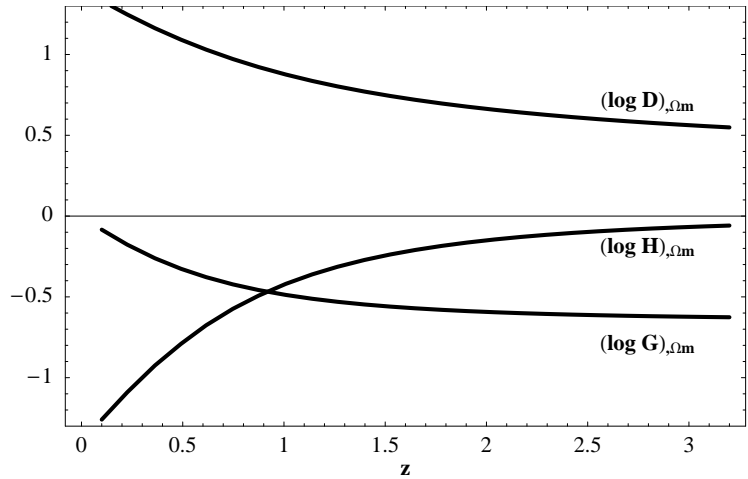}

\caption{Derivatives $\partial \log X/\partial \Omega _{m0}$ as function
of $z$, where $X=H,D,G$. }
\end{figure}

Unless differently specified, the reference cosmology is $\Lambda CDM$
with the same parameters as in SE, $\Omega _{m}=0.35,h=0.65,\omega _{b}=0.021,\tau =0.05,n_{s}=1,T/S=0,P_{s}=0$
.

\section{The surveys}

One of the main aim of the paper is to show how it is possible to
build, using deep and wide multicolor surveys, a photometric redshift
sample of galaxies suitable for the analysis of the galaxy power spectrum
as a function of redshift. Previous analyses reached two main conclusions
about the possible strategy. First of all spectroscopic surveys are
clearly favorite with respect to photometric surveys because any uncertainty
$>1$\% in the photometric redshifts increases the uncertainty on
$H(z)$ at $z>0.5$ to $>10$\%, for surveys covering areas of $100-1000$
deg$^{2}$. Second, the relatively bright magnitude limit needed to
keep a high completeness level in the spectroscopic follow up implies
a low number density of sources per unit area. Large survey areas
are hence needed to guarantee adequate total number of objects. Here
we show that it is possible to envisage a photometric redshift survey
with adequate depth and limited area of the sky at $z=0.5-1$ and
at $z=2.7-3.5$ that can provide constraints on the redshift dependence
of the DE equation of state which are competitive with those obtained
from SNIa and CMB in the foreseeable future. 

To measure the redshift evolution of $H(z)$ and of the angular diameter
it is important to analyse the power spectrum in a wide redshift interval
with specific sampling at $z=0.5-1$ and especially at $z=3$ where
the linear regime of galaxy clustering extends to smaller physical
scales allowing the analysis of acoustic oscillations to higher frequencies
in the range $0.1<k<0.5$ $h$ Mpc$^{-1}$. What is important is the
estimate of the characteristic wavelength of the acoustic oscillations,
$k_{A}=2\pi /s$ which is connected with sound horizon $s$ at recombination.
For our standard $\Lambda $CDM model $k_{A}\simeq 0.06$ $h$ Mpc$^{-1}$.
It is important to emphasize that local surveys could probe only the
first peak at $k\simeq 0.075$ $h$ Mpc$^{-1}$. Thus, an appropriate
comparison with the CMB power spectrum requires higher frequencies
and consequently high redshift surveys, in particular at $z=3$. 

Indeed, large scale galaxy surveys at $z\geq 3$ could in principle
be built in a straightforward way using the standard multicolor dropout
technique which exploits the detection in the galaxy spectra of the
UV Lyman absorption due to the local ISM in the galaxy and in the
intergalactic medium present along the line of sight. At $z=2.5-3.5$
the Lyman series absorption progressively enters the UV filter causing
a reddening in the U-R galaxy color. At $z=3.5-4.5$ the dropout shift
in the blue causing a reddening in the B-R galaxy color and so on.
Spectroscopic confirmation has provided a relatively high success
rate 75\% (Steidel et al. 1996, 1998) although in practice it is difficult
to obtain a high completeness level in the magnitude interval $R=24-25.5$,
which is the present limit for systematic spectroscopic follow up
at 10m class telescopes. A survey completeness level of 90\% could
be reached more easily at $R<24.5$ but with a surface density of
$\sim 0.3$ arcmin$^{-2}$ at $z=2.5-3.5$. The color selection of
the sample requires very deep exposures in the UV band implying efficient
UV imagers at 8-10m class telescopes to cover a large area in a short
time. Surveys at $z>3.5$ which use a B-dropout technique can avoid
high UV efficiency but the surface density of objects at the spectroscopic
limit of $I\sim 25$ is only $\sim 0.1$ arcmin$^{-2}$. In other
words galaxies at $z\sim 4-5$ are rarer and fainter. With these considerations,
surveys at $z\sim 3$ down to a magnitude limit of $R\sim 25$ are
the best compromise for the study of large scale structures at high
redshift provided that efficient wide field imagers with high UV efficiency
are used. The Large Binocular Camera in construction at the prime
focus of the Large Binocular Telescope is an ideal imager to this
aim (Pedichini et al. 2003). 

It is well known from the previous studies that there are two sources
of statistical error in the power spectrum estimate. The sample variance,
due to the number of wavelength bins, depends on the survey volume
and on the scale of the selected acoustic peak in the linear regime
(e.g. Peacock \& West 1992; Blake \& Glazebrook 2003): 

\[
({\frac{\sigma _{P}}{P}})^{2}=2\times {\frac{(2\pi )^{3}}{V}}\times {\frac{1}{4\pi k^{2}\Delta k}}\,.\]

The shot noise is the other source of noise. For a fixed volume it
decreases with increasing $nP>1$, where $n$ is the comoving galaxy
number density. In practice it is possible to show that the decrease
of the shot noise saturates for $nP>>1$ and values in the range $1<nP<5$
are more appropriate (SE). 

To reach at least the second peak at $k=0.135$ h Mpc$^{-1}$ it has
been shown that adopting $\Delta k=k_{A}/4=0.015$ h Mpc$^{-1}$ to
ensure an adequate wavelength sampling, a value $\frac{\sigma _{P}}{P}=0.02$
is needed as the fractional height of the peak is about 4\% (Blake
\& Glazebrook 2003). This requires that volumes of the order of $3.6\times 10^{8}$
$h^{-3}$ Mpc$^{3}$ should be mapped at $z=0.5-1$ and at $z=3$.
Since the volume per square degree is $V\simeq 3.5\times 10^{6}$
h$^{-3}$ Mpc$^{3}$ deg$^{-2}$ at $z=2.5-3.5$ and $V=1.7\times 10^{6}$
Mpc$^{3}$ deg$^{-2}$ at $z=0.5-1.3$ this implies surveys at least
of the order of $100-200$ deg$^{2}$ at $z\sim 3$ and $z\sim 1$,
respectively. 

Concerning the shot noise, the condition of $1<nP<5$ implies $n\sim 4-20\times 10^{-4}\sigma _{8,g}^{2} h^{3}$
Mpc$^{-3}$ respectively, given $P\approx 2500\sigma _{8,g}^{2}$
(Mpc/$h$)$^{3}$ at $k=0.2$ $h$/Mpc. 

At $z=3$, a photometric redshift survey down to the magnitude limit
of $R\leq 25$ gives a total number density in the range $z=2.7-3.5$
of $n\sim 2\times 10^{-3}$ $h^{3}$ Mpc$^{-3}$ after integration
over the UV Schechter luminosity function of Poli et al. (2001). 

At $z=1$, the same photometric survey at $R\leq 25$ in the range
$z=0.5-1$ would give a total number density of $\sim 4.5\times 10^{-2}$
$h^{3}$ Mpc$^{-3}$ as obtained from the blue luminosity function
of Poli et al. (2003). For sake of comparison, we adopt the same redshift
binning at $z\approx 1$ as SE.

Detailed densities and volumes for the basic surveys of area $100$deg$^{2}$
are shown in Table II in the same redshift intervals used by SE (except
we restrict conservatively the deepest survey to $z>2.7$ instead
of $z>2.5$). We will present results also for larger surveys, corresponding
to areas of 200 deg$^{2}$ and 1000 deg$^{2}$ respectively, and with
sparser surveys at $z\approx 1$, adopting a density 10 times smaller
as that quoted in Table II. The absolute error in redshift is taken
to be either zero (spectroscopic survey) or $\delta z=0.02$ and $\delta z=0.04$
(photometric surveys). Finally, we also include results for a \emph{hybrid}
survey, i.e. a photometric survey with $\delta z=0.02$ at $z\approx 1$
combined with a spectroscopic at $z\approx 3$. This choice depends
on the fact that multi object spectroscopy of a complete color selected
sample of Lyman break galaxies at $z=3$ is able to give more stringent
constraints on the cosmological parameters (see Fig. 3) than a similar
survey at $z\approx 1$ in the same area. Moreover a spectroscopic
survey at $z\approx 3$ appears also more efficient in terms of multiplexing
and completeness level if the magnitude limit is $R\leq 25$, corresponding
to a surface density $\leq 1$ arcmin$^{2}$. At $z=1$, on the contrary,
the average surface density even at the relatively shallow limit $R=23$
is already of the order of 8 arcmin$^{2}$ implying a multiplexing
of more than 460 targets in a field of e.g. $7\times 8$ arcmin$^{2}$
to be compared with the multiplexing capabilities of recent instruments
at 8m class telescopes like VIMOS at ESO/VLT where only 150 targets
can be observed in a single exposure in the same area.

\vspace{.2in}

\begin{center}\begin{tabular}{|c|c|c|}
\hline 
\multicolumn{3}{|c|}{Table II. Surveys}\\
\hline
$z$&
 $V_{s}$ (Gpc$^{3}$)&
 $n$\\
\hline
$0.5-0.7$&
 0.030&
 $6.9\cdot 10^{-2}$\\
\hline
$0.7-0.9$&
 0.040&
 $4.2\cdot 10^{-2}$\\
\hline
$0.9-1.1$&
 0.050&
 $3.1\cdot 10^{-2}$\\
\hline
$1.1-1.3$&
 0.057&
 $2.4\cdot 10^{-2}$\\
\hline
$2.7-3.5$&
 0.27&
 $2\cdot 10^{-3}$\\
\hline
\end{tabular}\end{center}

\vspace{.2in}

As in SE, to these surveys we add data from SDSS (Eisenstein et al.
2001). The SDSS data is always considered spectroscopic.

Redshift errors of the order of 1-2\% from photometric surveys are
demanding but still reachable if the photometric accuracy is $\delta m\simeq 0.02$
and if a good wavelength coverage and sampling is guaranteed (see
e.g. Wolf et al. 2001). This requires broad and intermediate band
imaging on wide field imagers at 8m class telescopes like Suprim-Cam
at the SUBARU or the upcoming Large Binocular Camera at the Large
Binocular Telescope (LBT). In particular the use of intermediate band
filters in the optical range can be important to ensure the required
redshift accuracy in the specific redshift intervals $z\sim 1$ and
$z\sim 3$. In the redshift interval $z=1.5-2.5$ near infrared imaging
is needed on wide areas but this requires a new generation of NIR
instruments.

To the redshift surveys we add CMB data, adopting the same Planck-like
specifications of SE and EHT. Finally, we evaluate SN constraints
from surveys that reproduce approximatively near future catalogs from
ground-based observations: 400 SNIa distributed uniformly in redshift
between $z=0$ and $z=1.5$, with magnitude errors $\Delta m=0.25$
and the same stretch factor correction as in the sample of Perlmutter
et al. (1999). We marginalize over the $H$-dependent magnitude $\mathcal{M}$
and over the stretch factor. These SN likelihood functions are therefore
completely independent of the Hubble constant determination. We also
compare with SN likelihood contours obtained assuming an independent
10\% gaussian error on $H_{0}.$ Notice that the SN likelihood contours
are estimated through the full likelihood function rather than with
a Fisher matrix  to facilitate the comparison with current estimates. 
 We will compare the LSS forecasts also with the present
SN dataset of Riess et al. (2004). 

For the bias of the galaxy populations at $z=1$ and $z=3$ we adopt
the same values as in SE, i.e. $b\sim 1-1.5$ at $z\sim 0.5-1$ and
$b\sim 3.3$ at $z\approx 3$ (see Adelberger et al. 1998). Specifically,
we assume $\sigma _{8,m}(z=0)=0.9$ and $\sigma _{8,m}(z)=\sigma _{8,m}(0)G(z)$
and $\sigma _{8,g}=1$ at $z=3$ and $\sigma _{8,g}=1.8$ at all other
redshift bins.

\section{Perfect fluid dark energy}

To review notation and to check our code against the results of SE
we begin the analysis with their DE model, based on a perfect fluid
with equation of state $p=w(z)\rho $ where\begin{equation}
w(z)=w_{0}+w_{1}z\,.\end{equation}
 Notice that such a perfect fluid model can be only a phenomenological
description of the background dynamics at low $z$. Beside the obvious
fact that $w$ diverges at high $z$, the problem is that a meaningful
derivation of the fluctuation equations  requires as additional
prescription the dependence of $w$ on the perturbed dark energy density.
Moreover, if one assumes for simplicity that $\delta w/\delta \rho =0$,
then dark energy becomes highly unstable at small scales since the
sound speed $c_{s}\equiv \sqrt{\delta p/\delta \rho }$ is imaginary
for $w(z)<0$. We can discuss therefore the growth function only in
the $\Lambda $CDM limit $w_{0}=-1,w_{1}=0$. In this limit a useful
approximation is\begin{equation}
\frac{d\log G}{d\log a}=\Omega _{m}(a)^{\gamma }\label{eq:glaw}\,,\end{equation}
 with $\gamma \approx 0.6$. Since $G$ can be calculated only in
the $\Lambda $CDM limit, we assume that $G$ does not depend on $w_{0},w_{1}$:
this is a conservative statement, since any dependence would further
restrict the area of the confidence region. Moreover, we found that,
naively extending the validity of (\ref{eq:glaw}) to other values
of $w_{0},w_{1}$ (i.e. assuming dark energy to be completely homogeneous),
the results do not change sensitively. Therefore $G$ will be assumed
in this section to depend only on $\Omega _{m0}$. The derivative
with respect to $\Omega _{m0}$ is
\begin{equation}
\frac{\partial \log G}{\partial \Omega _{m0}}  =  \gamma \int \Omega _{m}(\alpha )^{\gamma }\frac{d\log \Omega _{m}}{d\Omega _{m0}}d\alpha =
  -\gamma \int \frac{e^{3\alpha }\Omega _{m}(\alpha )^{\gamma }}{[e^{3\alpha }(\Omega _{m0}-1)-\Omega _{m0}]\Omega _{m0}}d\alpha \,.
\end{equation}

Once we have the total Fisher matrix we invert it and extract a submatrix
$F_{sub,ij}$ containing all the parameters $p_{i}$ that depend on
the cosmological parameters $\omega_m,\Omega _{m0},w_{0},w_{1}$. Beside $\omega_m,\Omega_{m0}$ these are, {\it for each survey}, \begin{equation}
\log D,\log H,\log G
\,,\end{equation}
 plus $\log D$ for the CMB. The diagonal elements of $F_{sub,ij}^{-1}$
give the square of the errors on the relative variables. The relative
errors on $H,D,G$ are plotted in Fig. 2 for the various surveys and
redshift errors. These intermediate results are consistent with those
given in SE (who however considered a different combination of surveys
and volumes). As remarked in SE, the errors on $D$ (contrary to $H$
and $G$) depend only weakly on redshift uncertainty. At $z=1$ the
angular diameter distance can be estimated to 4.3 \% in a spectroscopic
survey of area 200 deg$^{2}$ and to 7.5\%(10.0\%) in a similar survey
with $\delta z=0.02(0.04)$. Roughly a factor of 2 is gained by extending
to 1000 $\deg ^{2}$.

Changing to the DE variables \begin{equation}
d_{n}=(\Omega _{m0}h^{2},\Omega _{m0},w_{0},w_{1})\,,\end{equation}
 we obtain finally the projected DE Fisher matrix\begin{equation}
F_{DE,mn}=\left(\frac{\partial p_{i}}{\partial d_{m}}\right)F_{sub,ij}\left(\frac{\partial p_{j}}{\partial d_{n}}\right)\,.\end{equation}
 Then we again invert and take a submatrix $M_{w_{0},w_{1}}$ whose
eigenvalues and eigenvectors define the confidence ellipsoid on the
plane $w_{0},w_{1}$. Notice that a two-dimensional confidence region
corresponding to 1$\sigma $ includes only 39\% of likelihood. Here
and in all subsequent plots the confidence regions are given to 68\%
: this requires therefore that the 1$\sigma $ values of the parameters
(i.e. the eigenvalues of the two-dimensional Fisher submatrix) are
multiplied by  1.51.

The results for the various surveys and for different redshift errors
are in Figs. 3-6. In Fig. 3 and 4 we show the effect of marginalizing
over $G$ versus fixing $G=G(\Omega _{m0})$ as in Eq. (\ref{eq:glaw})
for the four surveys at $z\approx 1$ and the survey at $z=3$, always
including CMB and SDSS. As it can be seen, the effect is rather strong,
especially at $z=3,$ where the information in $G$ almost halves
the errors in $w_{0},w_{1}$. At this redshift, we could say that
a survey of 200 $\deg ^{2}$ that includes the information in $G$
(from now on, we call this approach {}``including $G$'') is roughly
equivalent in terms of constraining power to a survey of 1000 $\deg ^{2}$
with $G$ marginalization ({}``marginalizing over $G$''). The effect
at $z\approx 1$ is smaller, because $\partial \log G/\partial \Omega _{m0}$
is smaller at low $z$ (see Fig. 1).

In Fig. 5 we show the results for all redshift bins marginalizing
over $G$ as in SE while in Fig. 6 we include $G(z)$. In the same
plots we show the contours of the likelihood for simulated SNIa data.
These are obtained by a full likelihood analysis of simulated datasets,
marginalizing over the Hubble parameter derived from the SNIa themselves.
This renders SNIa sensitive to the slope $d\log D_{L}/dz$ of the
luminosity distance $D_{L}$, rather than to its absolute value. Since
\begin{equation}
\frac{d\log D_{L}}{dz}\propto \frac{1}{H(z)D_{L}(z)}\,,\end{equation}
 it is clear that the SNIa data is nearly degenerated in the direction
of $HD=const$, at $z\approx 1$. In Figs. 5-6 we also show the 68\%
confidence region for SNIa obtained assuming instead a 10\% gaussian
error in $H_{0}$. 

The $1\sigma $ errors on $\omega _{m},\Omega _{m},w_{0},w_{1}$ are
reported in Table III and IV, where we include always $G(z)$. For $\omega _{m}$
we give the relative errors; for the other parameters the absolute
one. Let us consider first the \emph{spectroscopic} results. In the
best case (all redshift bins over an area of 1000 $\deg ^{2}$ survey)
the errors on $w_{0}$ and $w_{1}$ are 0.12 and 0.16, respectively.
Marginalizing over $G$ we obtain 0.19 and 0.23: the information in
$G$ reduces the errors by roughly 30\% . The error on $\Omega _{m}$
goes from $0.017$ including $G$ to $0.03$ marginalizing over $G$.

For the \emph{photometric} case ($\delta z=0.02$) the best case of Table IV gives
$\sigma _{w_{0}}=0.32$ and $\sigma _{w_{1}}=0.39$. A hybrid survey
could reduce the errors to $0.23$ and 0.23, respectively. Reducing
the density of the $z\approx 1$ surveys leaves almost unaltered the
constraints in the spectroscopic case (since the ratio $nP/(1+nP)$
in Eq. (\ref{eq:stn}) is already close to saturation) while weakens
the photometric constraints by a few percent (see Table IV).

The relative error on the Hubble parameter $h=(\omega_m/\Omega_{m0})^{1/2}$ is 
dominated by  the relative error on $\Omega_{m0}$ and is therefore approximated by 
$ \sigma_{\Omega_{m0}}/2\Omega_{m0}$. We find $\Delta h/h\approx 3-7\%$.

It is interesting to comment also on the inclusion of SNIa data. The
supernovae constrain $w_{0}$ to a much higher precision than $w_{1}$.
Multiplying the baryon oscillations contours with the SN limits one
sees that the advantage of spectroscopy reduces considerably, at least
below 200 $\deg ^{2}$, almost regardless of $\delta z$, as can be
seen in Fig. 7.

We compare our baryon oscillations forecasts also with the \emph{present}
SN data. In Fig. 8 we show the constraints from the current set of
supernovae of Riess et al. (2004; \emph{gold} sample), along with
the constraints for a 200 $\deg ^{2}$ survey (marginalizing over
$G$) with reference cosmology centered on the Riess et al. (2004)
best fit, $w_{0}\approx -1.3,w_{1}\approx 1.5$ and $\Omega _{m0}=0.27$. This plot
shows that the size and orientation of the Fisher contour regions
depends in a significant way on the underlying reference cosmology,
as already remarked in SE. It also reveals that, in cases as this,
the spectroscopic surveys maintain their advantage even using the
SN information.

\vspace{.2in}

\begin{center}\begin{tabular}{|c|c|c|c|c|c|}
\hline 
\multicolumn{6}{|c|}{Table III. Spectroscopic surveys}\\
\hline
Area (deg$^{2}$)&
 $\sigma _{\omega _{m}}$&
 $\sigma _{\Omega _{m}}$&
 $\sigma _{w_{0}}$&
 $\sigma _{w_{1}}$&
\\
\hline
\multicolumn{6}{|c|}{$z=1$}\\
\hline
100&
 0.0053&
 0.037&
 0.39&
 0.56&
\\
\hline
200&
 0.0053&
 0.033&
 0.34&
 0.48&
\\
\hline
1000&
 0.0052&
 0.024&
 0.22&
 0.31&
\\
\hline
\multicolumn{6}{|c|}{$z=3$}\\
\hline
100&
 0.0051&
 0.045&
 0.32&
 0.36&
\\
\hline
200&
 0.0050&
 0.043&
 0.31&
 0.33&
\\
\hline
1000&
 0.0041&
 0.037&
 0.25&
 0.25&
\\
\hline
\multicolumn{6}{|c|}{all surveys}\\
\hline
100&
 0.0051&
 0.034&
 0.26&
 0.32&
\\
\hline
200&
 0.0049&
 0.029&
 0.21&
 0.26&
\\
\hline
1000&
 0.0040&
 0.017&
 0.12&
 0.16&
\\
\hline
&
&
&
&
&
 \\
\hline
\end{tabular}\end{center}

\vspace{.2in}

\begin{center}\begin{tabular}{|c|c|c|c|c|c|}
\hline 
\multicolumn{6}{|c|}{Table IV. Photometric ($\delta z=0.02$) surveys}\\
\hline
Area (deg$^{2}$)&
 $\sigma _{\omega _{m}}$&
 $\sigma _{\Omega _{m}}$&
 $\sigma _{w_{0}}$&
 $\sigma _{w_{1}}$&
\\
\hline
\multicolumn{6}{|c|}{$z=1$}\\
\hline
100&
 0.0054&
 0.046&
 0.45&
 0.72&
\\
\hline
200&
 0.0054&
 0.046&
 0.42&
 0.63&
\\
\hline
1000&
 0.0053&
 0.043&
 0.35&
 0.45&
\\
\hline
\multicolumn{6}{|c|}{$z=3$}\\
\hline
100&
 0.0053&
 0.046&
 0.45&
 0.70&
\\
\hline
200&
 0.0053&
 0.046&
 0.42&
 0.62&
\\
\hline
1000&
 0.0052&
 0.045&
 0.35&
 0.45&
\\
\hline
\multicolumn{6}{|c|}{all surveys}\\
\hline
100&
 0.0053&
 0.045&
 0.42&
 0.62&
\\
\hline
200&
 0.0053&
 0.044&
 0.39&
 0.54&
\\
\hline
1000&
 0.0050&
 0.039&
 0.32&
 0.39&
\\
\hline 
\multicolumn{6}{|c|}{ hybrid surveys}\\
\hline
100&
0.0052&
0.044&
0.31&
0.34&
\\
\hline
200&
0.0050&
0.042&
0.30&
0.31&
\\
\hline
1000&
0.0041&
0.033&
0.23&
0.23&
\\
\hline
\multicolumn{6}{|c|}{ density/10 at $z\approx 1$}\\
\hline
100&
0.0053&
0.045&
0.44&
0.67&
\\
\hline
200&
0.0053&
0.045&
0.41&
0.59&
\\
\hline
1000&
0.0051&
0.040&
0.34&
0.42&
\\
\hline
\end{tabular}\end{center}

\vspace{.2in}

\begin{figure}
\includegraphics[  bb=0bp 200bp 598bp 843bp,
  clip, scale=0.7]{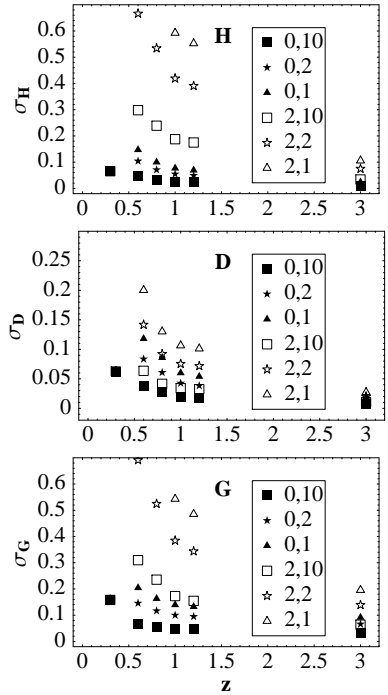}

\caption{Relative errors for $H,D,G$ for the various surveys. The legend
label $x,y$ means a survey with redshift error $x/100$ and
area $y\cdot 100$ $\deg ^{2}$. The dot at $z=0.3$ is the SDSS survey.}
\end{figure}

\begin{figure}
\includegraphics[  bb=100bp 300bp 598bp 843bp,
  clip, scale=0.7]{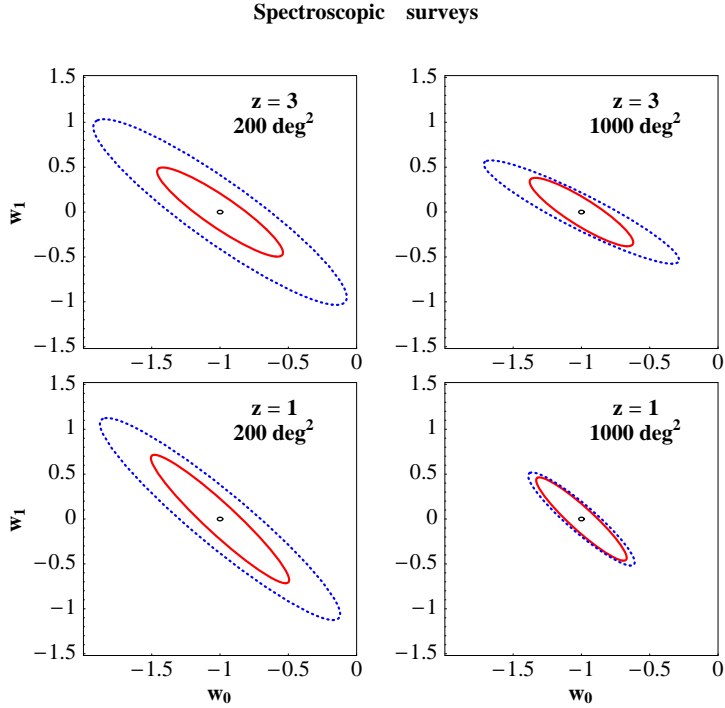}

\caption{Effect of the growth function for spectroscopic surveys: confidence
regions marginalizing over $G$ (dotted curves) and using the information
in $G$ (full curves). The plots are for the survey at $z=3$ alone
(upper panels) and for $z=1$ alone (lower panels). In the color version
of the figures lines referring to $G$-marginalization are blue, those
referring the the inclusion of $G$ are in red.}
\end{figure}

\begin{figure}
\includegraphics[  bb=100bp 400bp 598bp 843bp,
  clip, scale=0.7]{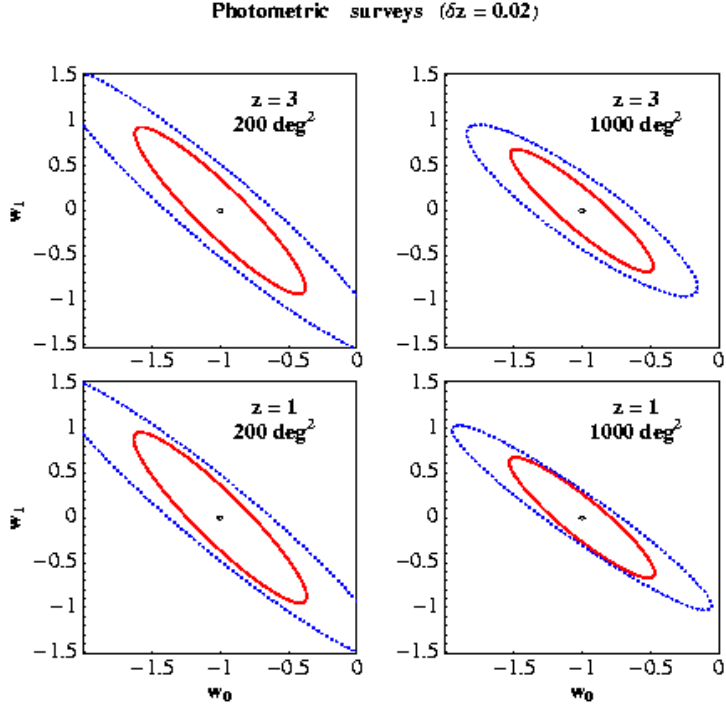}

\caption{Effect of the growth function for photometric surveys: confidence
regions marginalizing over $G$ (dotted curves) and including $G$
(full curves). The plots are for the survey at $z=3$ alone (upper
panels) and for $z=1$ alone (lower panels).}
\end{figure}
\begin{figure}
\includegraphics[  bb=100bp 300bp 598bp 843bp,
  clip, scale=0.7]{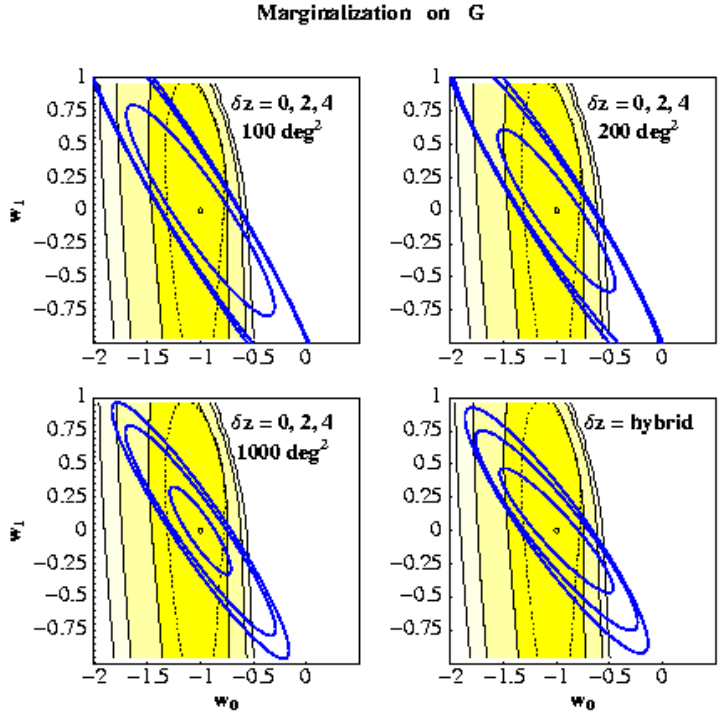}

\caption{Confidence ellipsis for the combined data CMB,SDSS, surveys at $z\approx 1,3$,
for three redshift errors, $0,0.02$ and $0.04$, inside to outside.
 Here and in all subsequent plots the redshift errors are indicated
as $100\cdot \delta z$. The central dot is the reference cosmology.
The bottom right panel refers to the hybrid survey discussed in the
text (spectroscopic plus photometric): here the contours are for surveys
of 1000, 200 and 100 $\deg ^{2}$, inside to outside. The shaded areas
are the confidence regions for 400 simulated SNIa distributed according
to the reference cosmology up to $z=1.5$, at 68\%, 95\% and 99\%
c.l., inside to outside, marginalizing over $H_{0}$ (uniform prior)
and over $\Omega _{m0}$ with a Gaussian centered on the reference
value and variance 0.1. The dotted contour (at $68\%$ c.l.) assumes instead a 10\%
Gaussian error on $H_{0}$ for the SNIa.}
\end{figure}

\begin{figure}
\includegraphics[  bb=100bp 300bp 598bp 843bp,
  clip, scale=0.7]{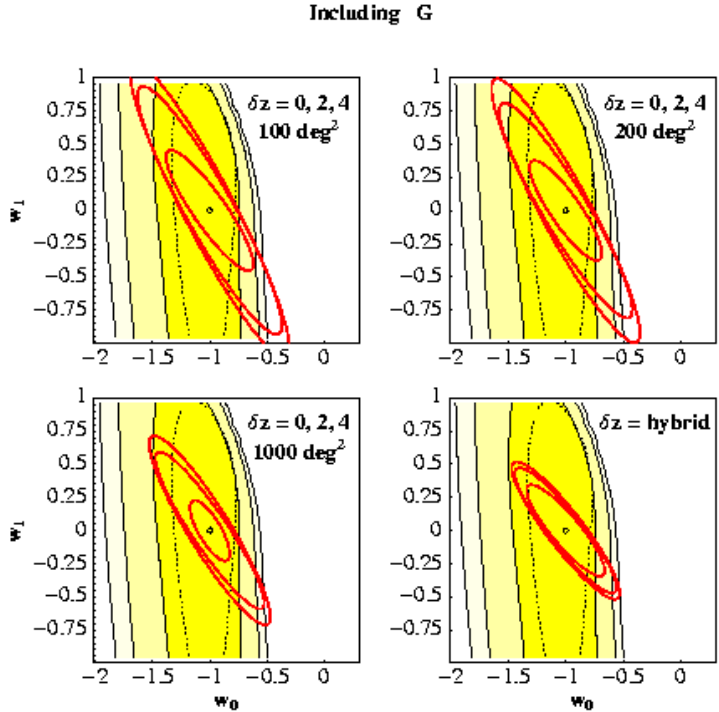}

\caption{As in Fig. 5 but including the dependence of the growth factor $G$
on the DE parameters.}
\end{figure}

\begin{figure}
\includegraphics[  bb=0bp 400bp 598bp 843bp,
  clip, scale=0.7]{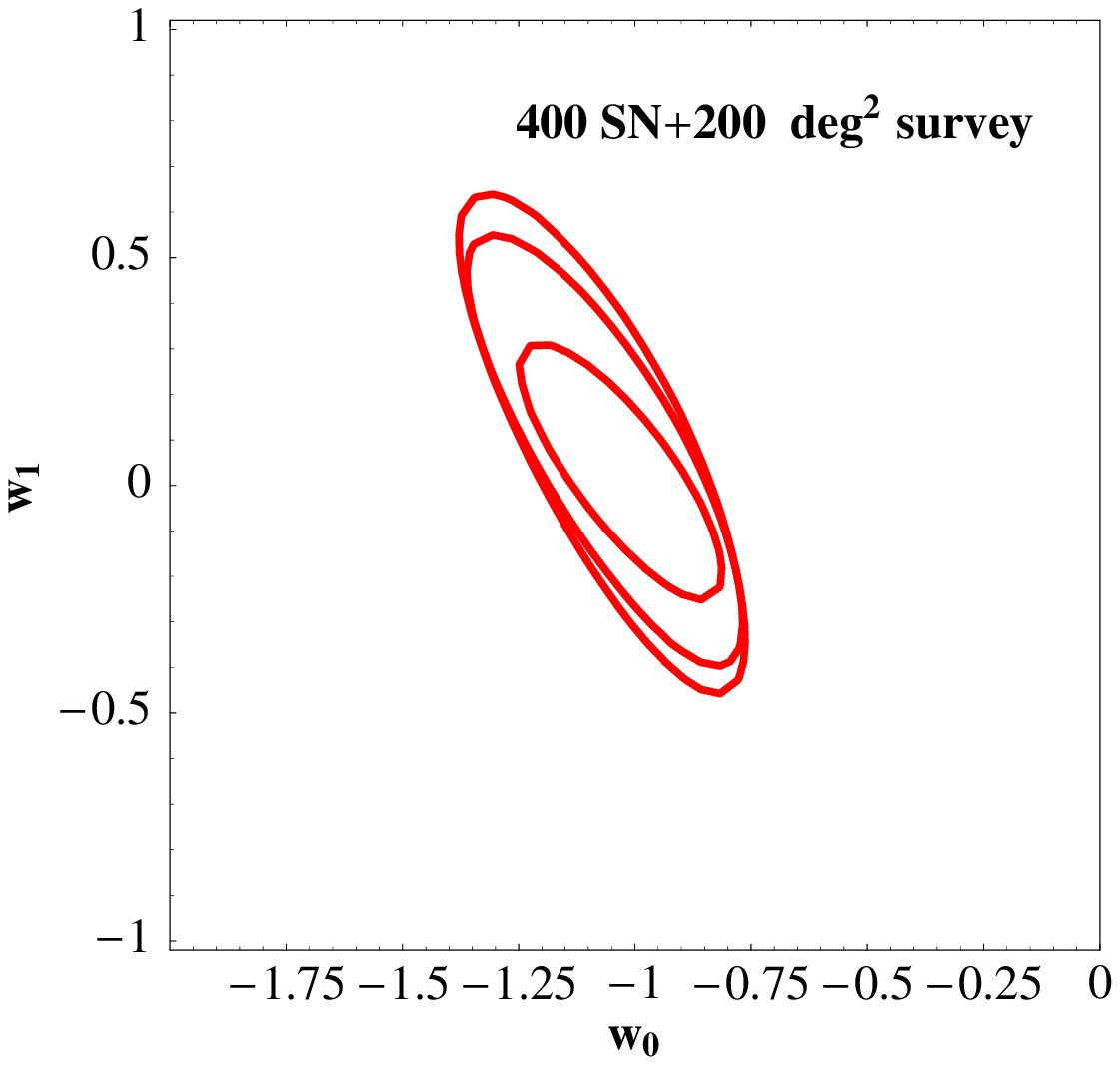}

\caption{Contours at 68\% of the combined likelihood of supernovae and surveys
of 200 $\deg ^{2}$ with redshift errors $\delta z=0,0.02,0.04$,
inside to outside (including $G$).}
\end{figure}

\begin{figure}
\includegraphics[  bb=0bp 400bp 598bp 843bp,
  clip, scale=0.5]{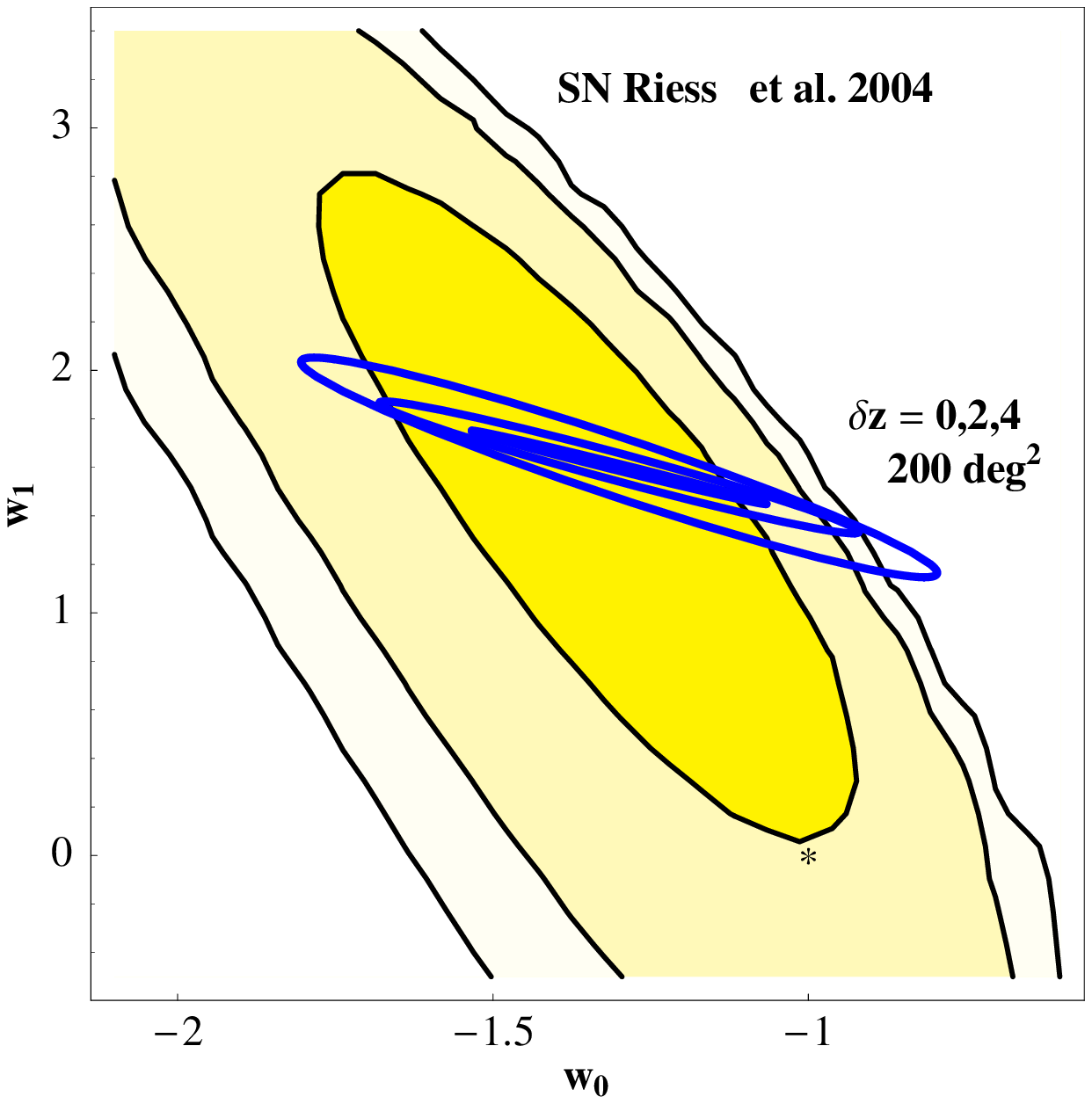}

\caption{Here we compare the confidence regions of 200$\deg ^{2}$ surveys
with redshift errors of $0,0.02$ and $0.04$ (marginalized over $G$)
to the recent SN data of Riess et al. (2004) assuming $\Omega _{m0}=0.27\pm 0.04$
(Gaussian errors). For this plot the reference cosmology in the Fisher
matrix coincides with the best fit to the SN ($w_{0}=-1.3,w_{1}=-1.5$)
. The SN contours are 68,90 and 95\% . The star marks the $\Lambda $CDM
model.}
\end{figure}

\section{Scalar field model}

The perfect fluid DE model of the previous section is a useful toy
model but has no theoretical motivation and is incomplete in the sense
that it is only a description of the background dynamics at low redshifts.
In this section we perform the analysis by considering an example
of a microphysical description of DE based on a scalar field model.

We adopt to this purpose one of the first and simplest scalar field
model proposed in literature, the inverse power-law potential (Ratra
\& Peebles 1988, Steinhardt et al. 1999)\begin{equation}
V=A\phi ^{-n}\,.\end{equation}
 The cosmological equations for this model are conveniently written in
the $e$-folding time $\alpha =\log a=-\log (1+z)$ as (the prime
denotes derivation with respect to $\alpha $)\begin{eqnarray}
\phi ''+\phi '(2+\frac{H'}{H})+a^{2}V_{,\phi } & = & 0 \,,\\
H^{2} & = & H_{0}^{2}\Omega _{m0}a^{-3}+\frac{\kappa ^{2}}{3}(\frac{1}{2}\phi '^{2}H^{2}+V)\label{eq:heq}
\,,\end{eqnarray}
 where $\kappa ^{2}=8\pi G$, $H_{0}^{-1}=3000$Mpc, and $\Omega _{K,P}$
denote the kinetic and potential energy, respectively, of the scalar
field. It is convenient to introduce the notation $x^{2}=\frac{\kappa ^{2}\phi '^{2}}{6}=\Omega _{K}$,
$y^{2}=\frac{\kappa ^{2}V}{3H^{2}}=\Omega _{P}$ and $h=H/H_{0}$ (which should not be confused with  the
dimensionless Hubble parameter).
Eq. (\ref{eq:heq}) becomes then (Amendola 2000) \begin{equation}
h^{2}=\frac{\omega _{m0}a^{-3}}{1-x^{2}-y^{2}}\,.\label{eq:hxy}\end{equation}
 Then the equations can be written as a dynamical system as\begin{eqnarray}
x' & = & -\frac{3}{2}x(x^{2}-y^{2}-1)-\mu y^{2(1+n)/n}h^{2/n}\,,\nonumber \\
y' & = & \mu xy^{(2+n)/n}h^{2/n}+\frac{3}{2}y(1+x^{2}-y^{2})\,,\label{eq:sys}\\
h' & = & -\frac{h}{2}(3+3x^{2}-3y^{2})\,,\nonumber 
\end{eqnarray}
 where we introduced the dimensionless quantity\[
\mu =-\frac{n}{\sqrt{2}}(\frac{\kappa ^{2}}{3})^{-\frac{n+2}{2n}}A^{-1/n}H_{0}^{2/n}=-n\sqrt{\frac{3}{2}}(\kappa \phi _{0})^{-1}\Omega _{P0}^{-1/n}\,,\]
 where $\phi _{0}$ is the present value of the scalar field and $\Omega _{P0}$
its present potential density. The system can be solved numerically
to derive $x(z),y(z)$ for each value of the parameters $n,\Omega _{m0},\mu $
but this is computationally cumbersome and hides interesting features
of the solutions. In Steinhardt et al. (1999) it has been shown in
fact that such a model possesses a tracking solution, defined as a
trajectory on which many initial conditions converge, as shown in
Fig. 9. Once a solution reaches this trajectory its properties are
determined solely by the slope $n$ and the present value $\Omega _{m0}$.
As long as $x,y\ll 1$ the tracking trajectory has $w_{\phi }\approx -2/(2+n)$;
when later on  DE takes over, $y$ tends to unity and $w_{\phi }\rightarrow -1$.
Since the variation of $w_{\phi }$ is relatively small, to a first
approximation the inverse power-law model for $n\ll 1$ resembles
a perfect fluid with $w_{0}\approx -2/(2+n)$ and $w_{1}\approx 0$.

In the Appendix we derive a semi-analytical fast approximation to
the tracking trajectory that interpolates correctly between the two
limits of $w_{\phi }$. This approximation will be used in what follows;
it is found to be precise to better than 1\% on $\Omega _{m}(z)$
and to better than 5\% on $w_{\phi }(z)$ across all the relevant
range. Further, we find that for the growth function the same approximation
$d\log G(z)/d\alpha =\Omega _{m}^{\gamma }(\alpha )$ adopted for
$\Lambda $CDM remains valid, with deviation from the numerical result
less than 0.1\% in the range $z\in (0,3)$ putting $\gamma =0.575$
for $n\approx 0$ and $\gamma =0.625$ for $n\approx 0.5$. For all
practical purposes the dependence of $\gamma $ on $n$ can be neglected
and we put $\gamma = 0.6$ in the whole interesting range.

\vspace{.2in}

\begin{figure}
\includegraphics[  bb=60bp 500bp 598bp 843bp,
  clip, scale=0.7]{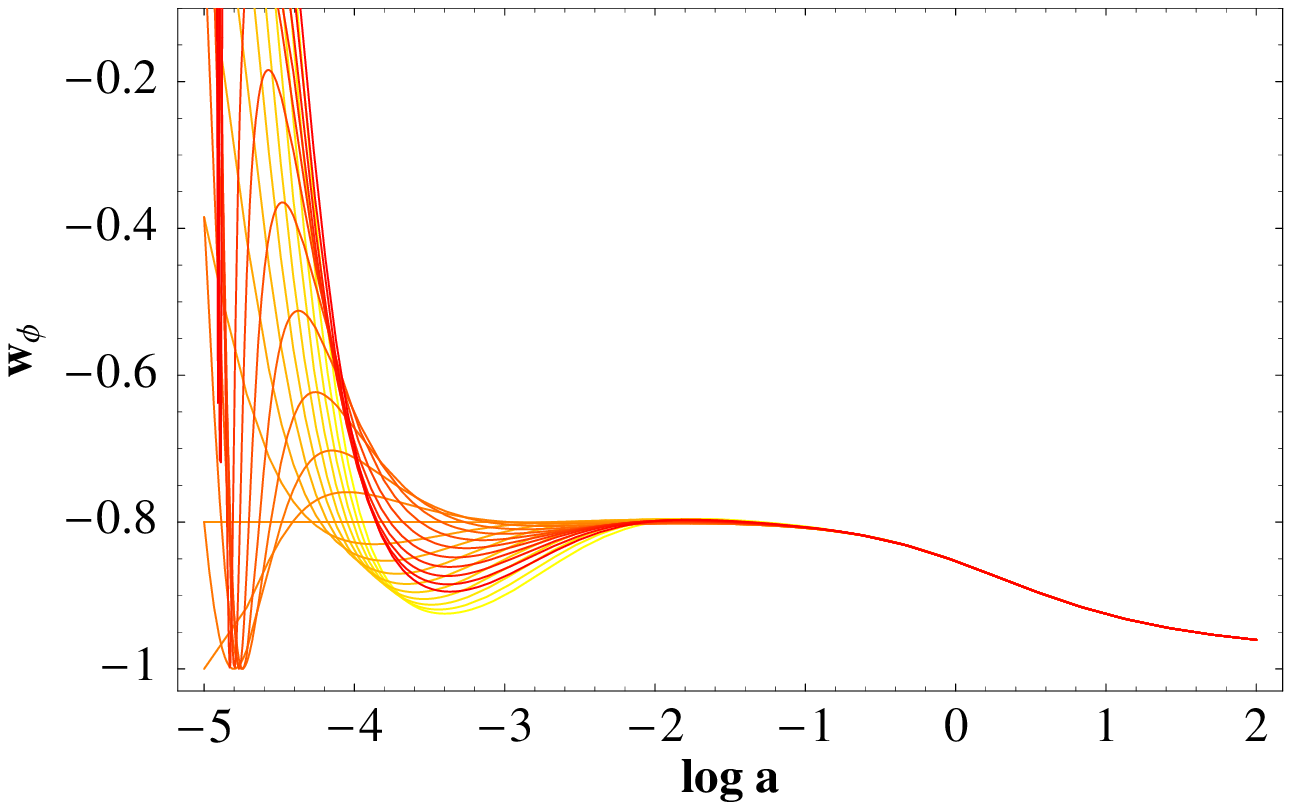}

\caption{Equation of state of the scalar field model for several
different initial conditions. The plot shows the tracking behavior:
all curves converge to a trajectory which, at early times, has constant
equation of state $w_{\phi }\approx -\frac{2}{2+n}$ and asymptotically
converge to $w_{\phi }=-1$. Here $n=0.5$ and $\Omega _{m0}=0.35$.}
\end{figure}

We project now the Fisher matrix in the subspace \begin{equation}
d_{n}=(\Omega _{m0}h^{2},\Omega _{m0},n)\,.\end{equation}

The results are shown in Fig. 10 (reference cosmology $n=0.05$, practically
indistinguishable from $\Lambda $CDM ) projected on the plane $\Omega _{m0},n$,
and in Table V. As expected, the SN contours are elongated towards
the contours of $HD$ (see the contour plots in Fig. 11). The contours
of the Fisher matrix are instead slanted towards the direction of
the iso-$G$ lines: this shows that it is the information in $G$
that drives the results. This optimizes the complementarity between
SNIa and LSS.

The tracking scalar field model has been tested towards SNIa and CMB.
We analysed the recent Riess et al. (2004) SNIa data with the inverse
power-law potential and found the constraint $n<0.9$ at 68\% c.l.
($n<0.5$ assuming $\Omega _{m0}=0.3\pm 0.05$). On the other hand
WMAP data allow for $n<0.94$ (or $n<0.74$ including the HST prior
on the Hubble constant), at 68\% c.l (Amendola \& Quercellini 2003).
From Fig. 10 and Table V we see that the baryon oscillations for a
scalar field model can constrain $\Omega _{m0}$ to within $0.01$
to $0.03$, and the slope $n$ to $n<$0.15 and $n<$0.53, depending
on the survey strategy. In all cases this is quite better than the
current and near-term SN and CMB constraint. Once again, most of the
advantage of spectroscopic measures vanishes if one takes into account
the SNIa contours (see Fig. 12)

Here again we perform the comparison with the Riess et al. (2004)
{\it gold} sample of supernovae (Fig. 13). It can be seen that the
present SN data already cut most of the vertical elongation of the
contours for the photometric surveys, leading to constraints on $n$
which are similar to those from spectroscopic surveys. The spectroscopic
errors on $\Omega _{m0}$ remain however competitive.

\section{Conclusions}

Future deep and wide redshift surveys will have the chance to explore
the little known landscape at $z\approx 1-3$ between the recent and early universe.
This is the epoch at which dark energy begins to impel its thrust
to the expansion: it is therefore of extreme interest for cosmology
in order to gain hold of reliable data on its evolution. As it has
been shown by the recent releases of high redshift SN data (Tonry
et al. 2003, Riess et al. 2004) the use of standard candles will have
much to improve before it will be able to set stringent limits on
the dark energy equation of state evolution. Moreover, there are still
no solid proposals for standard candles visible at redshifts significantly
higher than $z\approx 1.5$. Other probes of the expansion dynamics
at larger distances (gamma-ray bursts, lensing) might be promising
but depend on unknown physics and modeling. 

The use of the baryon oscillations in the matter power spectrum is
on the contrary based on well-known phenomena that have already seen
a spectacular validation in CMB observations. Moreover, it can provide
a tomography of the universe expansion from $z=0$ down to $z\approx 3.5$,
virtually without loss of precision. An intrinsic advantage of this
method is the combined use of $H,D$ and $G$ to reduce the degeneracies
of each of these quantity. We have shown that the inclusion of $G$
reduce the errors by 30\% roughly. We found that a 200$^{2}$ survey
with absolute redshift error of $\delta z=0.02$ can limit $w_{0},w_{1}$
by 0.39 and 0.54, respectively, assuming $w_{0}=-1,w_{1}=0$ as reference
values. A spectrocopic survey can reduce the uncertainties to 0.21
and 0.26. If the dark energy is modeled as a scalar field with inverse
power-law potential $n$, the limits are $n<0.26$ (spectroscopic
redshift) and $n<0.40$ (redshift error $\delta z=0.02$), along with
an estimation of $\Omega _{m0}$ to better than 3\%. Inclusion of
supernovae data further reduce the errorbars due to the different
direction of degeneracy. The reduction is found to be particularly efficient
in the case of photometric redshift measurements.

\begin{figure}
\includegraphics[  bb=100bp 300bp 598bp 843bp,
  clip, scale=0.7]{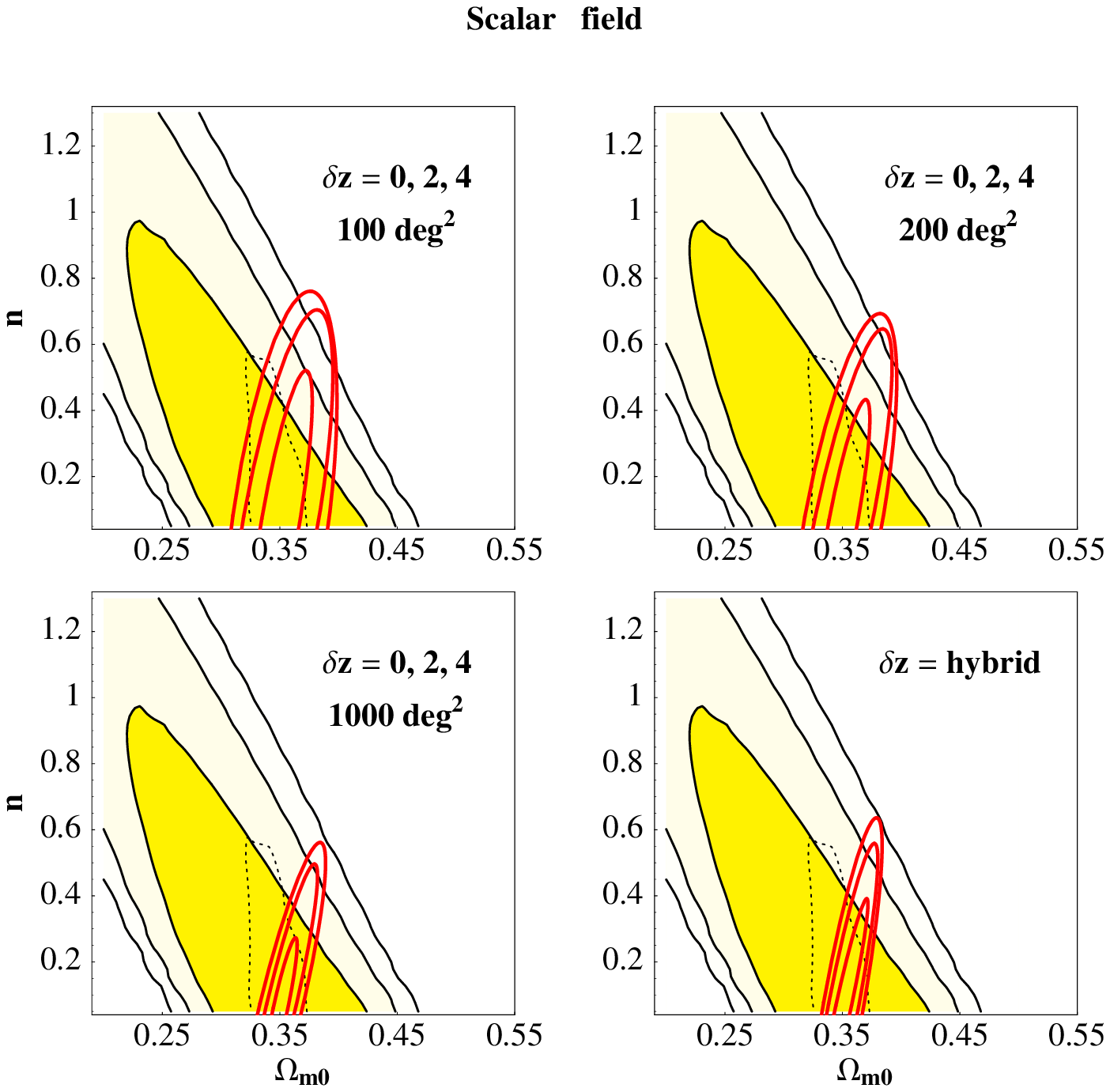}

\caption{As in Fig. 5 but for the scalar field model. Here the SNIa confidence
regions are not marginalized over $\Omega _{m0}$.}
\end{figure}

\begin{figure}
\includegraphics[  bb=0bp 100bp 598bp 843bp,
  clip, scale=0.7]{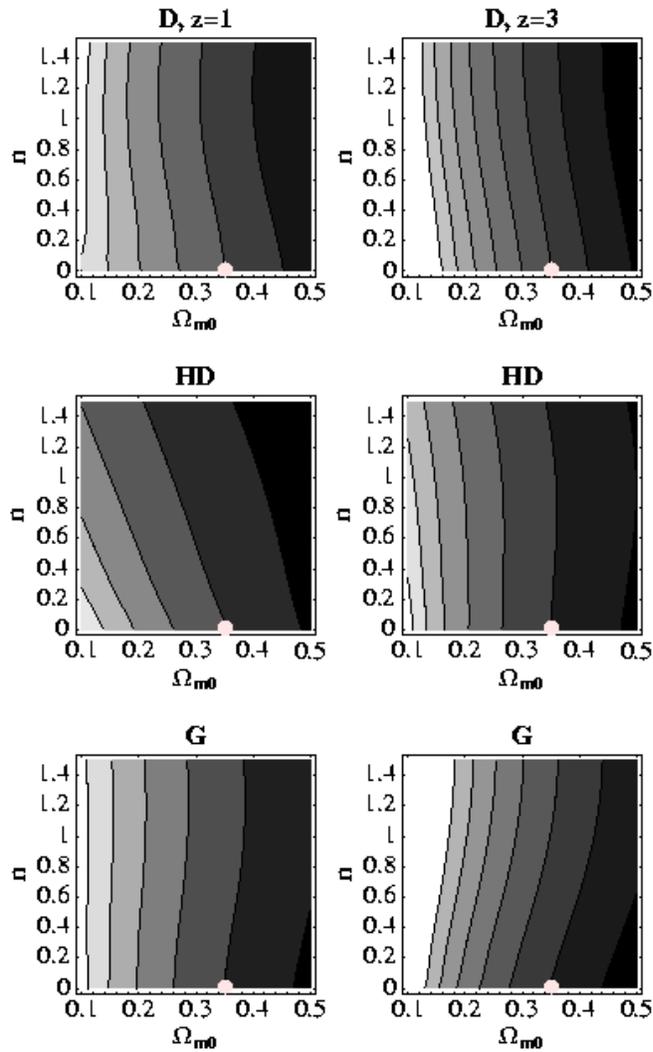}

\caption{The plot displays the contours of $D,HD,G$ at $z=1$ (left column)
and $z=3$ (right column). Each function is normalized by the value
at the reference cosmology, marked by the white dot. The contours
are for steps of 0.05, increasing from black to white}
\end{figure}

\begin{figure}
\includegraphics[  bb=0bp 400bp 598bp 843bp,
  clip, scale=0.7]{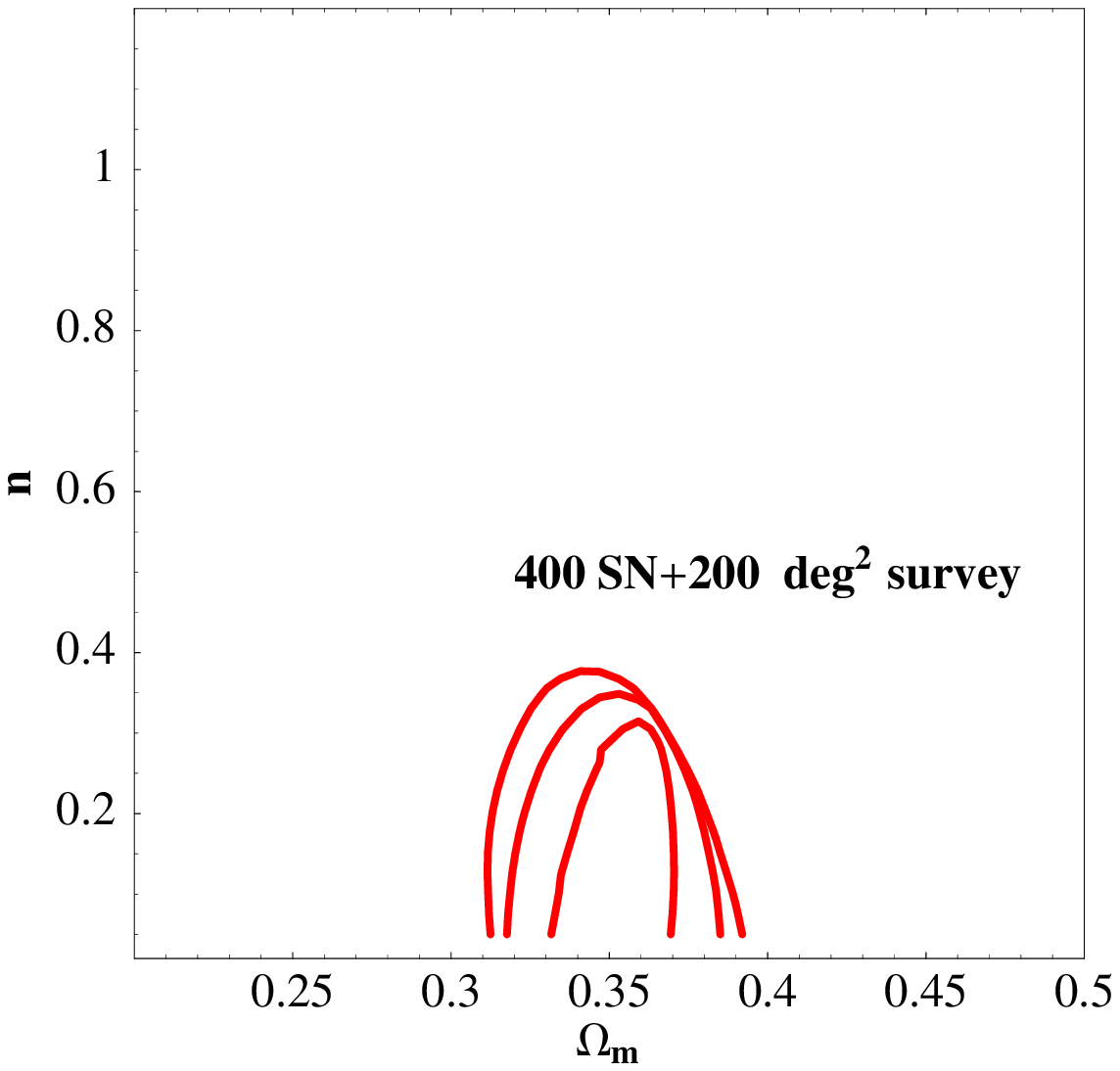}

\caption{Scalar field: contours at 68\% of the combined likelihood of supernovae and surveys
of 200 $\deg ^{2}$ with redshift errors $\delta z=0,0.02,0.04$,
inside to outside (including $G$).}
\end{figure}

\section*{Appendix. The fast-track approximation}

A tracking solution, as any other trajectory, is defined in a 3D phase
space by two relations of the form $e_{1,2}=f_{1,2}(x,y,h)$ where
$e_{1,2}$ are constants. There must exist therefore two constants
of motion of the  system (\ref{eq:sys}). In the original derivation of the tracking
solution (Steinhardt et al. 1999) it was found that the equation of state\begin{equation}
w=\frac{x^{2}-y^{2}}{x^{2}+y^{2}}\,,\label{eq:wdef}\end{equation}
 is almost constant during the tracking. The second constant of motion
is  \begin{equation}
q=y^{(2+n)/n}h^{2/n}\label{eq:qdef}\,.\end{equation}
 Then, by derivating, we obtain for the tracking solution\begin{eqnarray}
q_{t} & = & -\frac{3}{2\mu }\sqrt{1-w^{2}}\,,\nonumber \\
w_{t} & = & -\frac{2+n(y^{2}-x^2)}{n+2}\,.\label{eq:wgen}
\end{eqnarray}
 The second relation says that $w,q$ are constant only for $y^{2}-x^{2}\ll 1$:
this is the regime in which the tracking behavior was originally found.
In this case\begin{equation}
w=-\frac{2}{2+n }\label{eq:wtrack1}\,.\end{equation}
 However, since near the present epoch $x^{2},y^{2}$ are no longer
negligible, it is necessary to improve the accuracy over the solution
(\ref{eq:wtrack1}). Substituting $w$ from Eq. (\ref{eq:wdef}) in
(\ref{eq:wgen}) we obtain (assuming $w>-1$)

\begin{equation}
w(y,n)=\frac{1-2y^{2}-\sqrt{9-24(1+n)/n+16(1+n)^{2}/n^{2}-4y^{2}+4y^{4}}}{4/n+2}\,.\end{equation}
 From (\ref{eq:qdef}) we obtain\begin{equation}
y=(qh^{2/n})^{\frac{n}{2+n}}\,.\end{equation}
 Employing (\ref{eq:hxy}) we obtain finally\begin{equation}
y=\left[-\frac{3}{2\mu }\sqrt{1-w(y,n)^{2}}\left(\frac{\omega _{m}a^{-3}}{1-y^{2}(\frac{2}{1-w(y,n)})}\right)^{2/n}\right]^{\frac{n}{n+2}}\label{eq:ysol}\,.\end{equation}
 The parameter $\mu $ has to be fixed so that at the present $y^{2}+x^{2}=1-\Omega _{m0}$
. This gives a relation $\mu =\mu (n,\Omega _{m0})$. We found by
iterative integrations that a very good fit to $\mu $ is \begin{eqnarray*}
\mu  & = & \exp [-0.53-\frac{0.00036}{n^{3}}+\frac{0.0033+0.0088\Omega _{m0}}{n^{2}}-\\
 &  & \frac{0.13-1.16\Omega _{m0}+0.41\Omega _{m0}^{2}-1.69\Omega _{m0}^{3}}{n}+0.31n+2.43\Omega _{m0}-1.25\Omega _{m0}^{2}]\,.
\end{eqnarray*}
 With this substitution, Eq. (\ref{eq:ysol}) becomes an implicit
algebraic relation that, solved numerically for $y$, gives $y(a;n,\Omega _{m0})$
and all related quantities ($x(t),w(t),h(t)$).

\newpage
\vspace{.2in}
\begin{center}\begin{tabular}{|c|c|c|c|}
\hline 
\multicolumn{4}{|c|}{Table V. Results for scalar field.}\\
\hline
Area (deg$^{2}$)&
 $\sigma _{\omega _{m}}$&
 $\sigma _{\Omega _{m}}$&
 $\sigma _{n}$\\
\hline
\multicolumn{4}{|c|}{$\delta z=0$}\\
\hline
100&
 0.0032&
 0.018&
 0.32\\
\hline
200&
 0.030&
 0.016&
 0.26\\
\hline
1000&
 0.0022&
 0.010&
 0.15\\
\hline
\multicolumn{4}{|c|}{$\delta z=0.02$}\\
\hline
100&
 0.0034&
 0.030&
 0.44\\
\hline
200&
 0.0033&
 0.028&
 0.40\\
\hline
1000&
 0.0031&
 0.021&
 0.30\\
\hline
\multicolumn{4}{|c|}{$\delta z=0.04$}\\
\hline
100&
 0.0034&
 0.032&
 0.47\\
\hline
200&
 0.0034&
 0.030&
 0.43\\
\hline
1000&
 0.0033&
 0.026&
 0.34 \\
\hline
\multicolumn{4}{|c|}{hybrid surveys}\\
\hline
100&
0.0032&
0.023&
0.39\\
\hline
200&
0.0031&
0.020&
0.34\\
\hline
1000&
0.0026&
0.014&
0.23\\
\hline
\end{tabular}\end{center}

\newpage

\begin{figure}
\includegraphics[  bb=0bp 400bp 598bp 843bp,
  clip, scale=0.7]{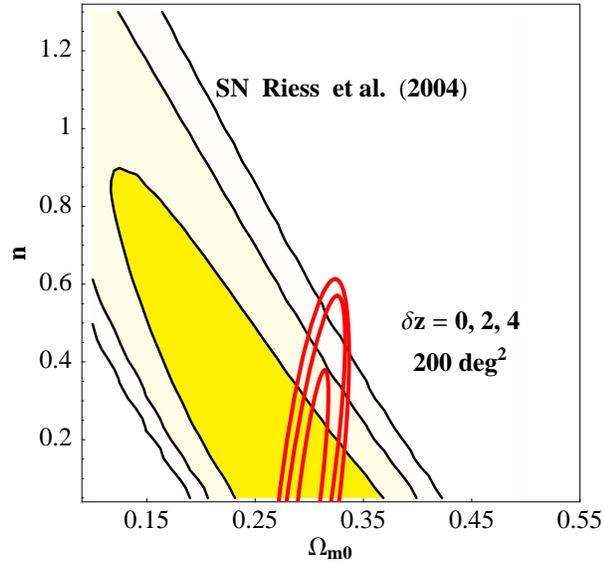}

\caption{Here we compare the confidence ellipsis of 200$\deg ^{2}$ surveys
with absolute redshift errors of $\delta z=0,$0.02 and $0.04$(inside
to outside), with the confidence regions for the scalar field model
with the SN data of Riess et al. (2004) For this plot the reference
cosmology in the Fisher matrix coincides with the best fit to the
SN ($\Omega _{m0}=0.3,n=0$). The SN contours are 68,90 and 95\% .}
\end{figure}

\label{lastpage}

\end{document}